\documentstyle[11pt]{article}
\input{psfig}
\makeatletter                                                           %LMS

\def\section{\@startsection {section}{1}{\z@}{-1.5ex plus -.5ex         %LMS
minus -.2ex}{1ex plus .2ex}{\large\bf}}                                 %LMS

\def\@thmcountersep{}                                                   %LMS

\long\def\@makecaption#1#2{\vskip 10pt \setbox\@tempboxa\hbox{#1. #2}   %LMS
   \ifdim \wd\@tempboxa >\hsize   % IF longer than one line:            %LMS
       #1. #2\par                 %   THEN set as ordinary paragraph.   %LMS
     \else                        %   ELSE  center.                     %LMS
       \hbox to\hsize{\hfil\box\@tempboxa\hfil}                         %LMS
   \fi}                                                                 %LMS

\def\ps@headings{                                                       %LMS
 \def\@oddhead{\footnotesize\rm\hfill\runninghead\hfill}               %LMS
 \def\@evenhead{\@oddhead}                                              %LMS
 \def\@oddfoot{\rm\hfill\thepage\hfill}\def\@evenfoot{\@oddfoot} }      %LMS

\def\@doubleleading{1.6}
\def\baselinestretch{\@doubleleading}

\makeatother                                                            %LMS
\title
{\bf On the Origin of Strong Magnetic Fields in Young Supernova Remnants}{}{}

\def\runninghead
{JUN and NORMAN:\quad  Strong Magnetic Fields in Young Supernova Remnants}

\author{\em Byung-Il Jun \thanks{
Present address : Department of Astronomy, University of Minnesota,
116 Church Street, S.E., Minneapolis, MN 55455, bjun@astro.spa.umn.edu}
 and Michael L. Norman \\
Laboratory for Computational Astrophysics \\ National
Center for Supercomputing Applications \\ Department of Astronomy \\
University of Illinois at Urbana-Champaign}
\begin{document}

\pagestyle{headings}                                                   %LMS
\flushbottom                                                           %LMS

\maketitle
\vspace{-10pt} % include according to taste.

\begin{abstract}
\noindent

 Young supernova remnants such as Tycho generally exhibit a bright 
circular clumpy shell in both radio and X-ray emission.  For several
young remnants, various arguments suggest that the magnetic field is
larger than can be explained by compression of a few $\mu G$ ambient magnetic
field by the shock wave.  Radio polarization studies
reveal a net radial orientation of magnetic fields in the shell which
cannot be explained by the simple compression either.
We model Rayleigh-Taylor instability at the interface of the ejecta and the
shocked ambient medium to explain these observations.  
We have performed multidimensional MHD simulations of the instability in the
shell of a Type-I supernova remnant for the first time utilizing
a moving grid technique which allows us to follow the growth of the
instability and its effect on the local magnetic field in detail.
We find that the evolution of the instability is very
sensitive to the deceleration of the ejecta and the evolutionary stage
of the remnant.
As the reverse shock enters the inner uniform density region,
Chevalier's self-similar stage ends and the thickness of radio
shell increases and the instability weakens.
Our simulation shows that Rayleigh-Taylor and Kelvin-Helmholtz 
instabilities amplify ambient magnetic fields locally by as much as a
factor of 60  around dense fingers due to stretching, winding,
and compression.   Globally, the amount of magnetic-field
amplification is low and the magnetic energy density reaches only
about 0.3\% of the turbulent energy density at the end of simulation.
Strong magnetic field lines draped around the
fingers produce the radial B-vector polarization, whereas thermal
bremsstrahlung from the dense fingers themselves produce the clumpy
X-ray emission.  As a result, the X-ray emission peaks inside of the
radio emission.   Surface brightness profile shows no detailed
correspondence between radio and X-ray emission. Major part of radio
and X-ray luminosity comes from the mixing region.

{\it Subject headings:} {Supernova remnants -- instability -- magnetic fields
-- MHD simulation}

\end{abstract}

\section{Introduction} % 1
\vspace{5mm}

   Extensive observations of young supernova remnants (hereafter SNRs) have
been made, especially for the prototypical sources Tycho, Kepler, and
Cas A.   These SNRs are believed to belong to the pre-Sedov stage
characterized by a shock expansion law that is faster than the
$t^{2/5}$ law expected for Sedov expansion (Tan and Gull 1985), and a
two-shock structure (Ardavan 1973; McKee 1974). 
These SNRs generally show a circular, clumpy shell near the outer
boundary of the remnant in the radio and X-ray emissions.  
The magnetic field strength
in the radio shell has been inferred to be in the range of $10^{-4}
\sim 10^{-3} G$ by assuming equipartition of energy between magnetic
fields and relativistic particles (Strom and Duin 1973; Henbest 1980;
Anderson et al 1991).  Other arguments suggest that the low limit of
the magnetic field in young SNRs is much larger than the compressed value
of ambient magnetic field obtained by simple shock compression (Cowsik
and Sarkar 1980; Matsui et al. 1984; Reynolds and Ellison 1992).
 The locations of the inner radio shell and the clumpy X-ray emission in
Tycho's SNR coincide although there is no
detailed correspondence between radio and X-ray features (Dickel et
al. 1991).  Detailed radio polarization
studies reveal that the magnetic field in the
shell exhibits a cellular pattern with a net radial orientation
(Downs and Thompson 1972; Milne 1987; Dickel et al. 1991).   
In addition to these generic features of young SNR, a thin bright
rim (sharp outer boundary) in radio emission exist in Tycho's SNR  
(Dickel et al 1991) and
the SNR of A.D. 1006 (Reynolds and Gilmore 1986,1993).  The rim in Tycho's
SNR has a typical degree of polarization of 20\% to 30\% while the
degree of polarization within a cell in the main shell is about 7\%
(Dickel et al 1991).   Tycho's SNR and the SNR of A.D. 1006 show 
the coincidence of sharp edges in radio and X-ray emission which is
generally interpreted as the locus of the shock front.  They are
presumed to have been Type I events implying that the remnant expanded
into a medium of roughly constant density.  Our study is motivated to
understand the origin of these generic features of young remnants of
Type I supernova by means of numerical simulation.

  It has been suggested that the contact interface
between the ejecta and the shocked ambient medium is Rayleigh-Taylor
(hereafter R-T) unstable (Gull 1973) due to the strong deceleration
felt by the denser ejecta as it sweeps up lighter ambient gas.  
The R-T instability in the
decelerating stage (pre-Sedov stage) has been studied
in one space dimension by Gull (1973,1975) and Dickel et
al. (1989). Chevalier et al. (1992) extended the study of the
instabilities to two dimensions by modeling the interaction region
assuming self-similar flows and without the inclusion of a magnetic
field.  They found that a turbulent mixing layer is produced in the
nonlinear state.  Such mixing could amplify the interstellar magnetic
field to the inferred milligauss values, and thus the incorporation of
magnetic fields into multidimensional simulations is of great importance.
Here, we present the first study of the evolution of the R-T instability in
young remnants of Type I supernova by two-dimensional MHD
simulations.   Since we simulate the entire remnant, our calculation 
is not restricted to the
self-similar stage, and thus we are able to study the growth and eventual
saturation of the instabilities for the evolving remnant.   As a
consequence, we are able to predict the evolution of the integrated
X-ray and radio luminosity of the remnant during the first 500 years.
Subsequent papers will extend this study to 3D MHD (Jun and Norman
1995b) and greater dynamical ages (Jun and Norman 1995c).

  The linear theory of the classical R-T instability has been fully
studied by Chandrasekhar (1961).  The nonlinear growth of the
instability has been studied numerically (Sharp 1984; Youngs
1984,1991).  In a previous work, we have explored the classical R-T 
instability in incompressible
magnetic fluids by two- and three-dimensional MHD simulations (Jun,
Norman, and Stone 1995, hereafter JNS).  
Although highly idealized, these simulations
support Gull's picture that the radial magnetic field in
the main shell of SNR can be produced as the R-T fingers stretch
existing field lines.   The present simulations put Gull's hypothesis
to the test in a realistic astrophysical context.
We find that Gull's model is able to explain a clumpy bright radio
shell which is generated by the turbulent amplification of magnetic
fields through the R-T instability.  Our simulation is also able to
explain radial magnetic fields in the main shell.  The evolution of the
instability and the radio shell is a sensitive function of the
deceleration of the ejecta and the evolutionary stage of the remnants.

 In this paper, we present the results of MHD simulations of the
instabilities in young SNRs in a uniform ambient medium and attempt
to explain some of observational features of Type I SNR.   
Section 2 presents the initial conditions and numerical
method.  Results are presented in Section 3, and a discussion of our
numerical limitations is given in Section 4.  We then compare our
model with observations and conclude in Section 5.

\vspace{5mm}
\section{Initial condition and numerical method} % 2
\vspace{5mm}

   We initialize the outer 3/7 of the star of mass 1.4 $ M_{\odot}$ with 
a power law density profile 
$\rho \sim r^{-7}$ and the inner 4/7 of the star with a constant
density to mimic Type-Ia supernova remnants (Colgate and McKee 1969).
An explosion of energy of $10^{51} ergs$ is deposited as 
kinetic energy where the velocity is assumed to be linearly proportional to
the radius.
The radius of the inner core ($r_c$), the density of the inner core
radius($\rho_c$), and the velocity at the outer boundary of the ejecta ($v_0$)
are now given by,
\begin{equation}
 r_c = R{(1 - {x(3-n)M \over 4\pi \rho_0 R^3})}^{1 \over 3-n},
\end{equation}
\begin{equation}
 \rho_c = {3(1-x)M \over 4\pi r_c^3},
\end{equation}
\begin{equation}
 v_0 = E_k^{1/2}{( {2\pi \rho_c r_c^5 \over 5R^2} + {2\pi \rho_0 R^3 (1
- {(R/r_c)}^{n-5}) \over 5-n})}^{-1/2}
\end{equation}
 where $R$ is the radius of the ejecta, $x$ is the mass fraction of the
outer power law density region which is 3/7 in our case, $n$ is the
density power index of the outer region, $M$ is the total mass of the
ejecta, $\rho_0$ is the density of ejecta at $R$, and $E_k$ is the
total kinetic energy.
The background density is chosen to be uniform
with $1.67 \times 10^{-24} g/c.c.$.   The temperature of the entire
region is taken as $10^{4} K$.
The background magnetic field is chosen to be $3.5 \mu G$ and 
assumed to be tangential to the shock front and lying in the
simulation plane.   A second
simulation assumed a toroidal magnetic field which is
perpendicular to the two-dimensional plane.  We will focus our study
on the case with an initial pure poloidal field since the amplification of the
magnetic field occurs by the stretching of the poloidal field line.
The density is 
initially perturbed within the entire region of computational space
with a random noise of 2.0\% amplitude.   We start the simulation 
at $R = 0.1 pc$.  Although this
value is larger than the stellar radius of the presupernova, the
important variable in the evolution is mass ratio rather than the
initial radius.  The mass ratio $\mu$ is defined by the swept-up mass by the
forward shock divided by the ejected mass.   This mass ratio represents
the dynamical age of the remnant because the
velocity of the remnant is determined by the mass ratio from
momentum conservation argument.   Therefore, the mass ratio is a
meaningful quantity for the age rather than time.

  We solve the ideal compressible MHD equations in a moving Eulerian  
grid using the
ZEUS-3D code originally written by David Clarke and further developed
by the Laboratory for Computational Astrophysics at the University of
Illinois, Urbana-Champaign.  Shocks are stabilized by von
Neumann-Richtmyer artificial viscosity.  Magnetic fields are
transported using Constrained Transport (Evans \& Hawley 1988)
modified with the Method of Characteristics (Stone \& Norman 1992b) to
satisfy the divergence-free constraint and to treat Alfven wave properly.
The code includes no physical diffusion term, and the numerical
diffusion is determined by the grid resolution.
Readers are referred to Stone \& Norman (1992 a,b) for the detailed
description of algorithms of the code.

 A moving grid method is used to maintain high resolution in the
region between the forward shock and the reverse shock.   
Basically, the velocity of the forward shock is measured
numerically every time step and it is assigned to the grid velocity at
the shock front.
All other grids are also expanding homologously at every timestep.
A large number of zones are distributed uniformly between the outer
and reverse shocks. The region inside the reverse shock is
resolved  with ratioed zones which increase in size as $r \to 0$ in
order to avoid a too stringent Courant condition on the timestep (Fig.1).
This method allows us to keep the high resolution at the intershock region
and to avoid a very small timestep near the origin.
We used a wedge of $\pi/4$ in spherical polar geometry ,and periodic
boundary conditions at the angular boundaries and reflecting boundary 
condition at the origin.   The 
intershock region is resolved with $200\times400$ grids and the region
inside the reverse shock is resolved with $100\times400$ grids.

 To track the contact discontinuity between the ejecta and the ambient
material, we evolve the mass fraction continuity equation (see JNS) :
\begin{equation}
{\partial (m_f \rho) \over \partial t} + {\partial (m_f \rho v_i)
\over \partial x_i} = 0 \nonumber
\end{equation}
where $m_f$ is the mass fraction of the ejecta.  The initial mass
fraction distribution is  1 for the ejecta and 0 for the ambient
medium.  This method is particulary useful in the study of the mixing layer.

\vspace{5mm}

\section{Results} % 3

\subsection{1D results} % 1
\vspace{5mm}
 
 We first carried out the simulation in one dimension to an age of 500
years to study the evolution of the remnant without the instability.
At this time, the mass ratio is 3.94.
The basic physical variables at 100 years ($\mu = 0.23$) are 
shown in Fig.2.   
It can be seen from the distribution of grid points that our moving grid
technique follows the forward shock accurately.
The intershock region agrees well with Chevalier's
self-similar solution (1982).  The magnetic field profile agrees well
with the simulation of Dickel et al.(1989).  
The magnetic field is the strongest at
the contact discontinuity at this time.   Nonetheless, the magnetic
field is essentially passive. This can be seen by comparing the
magnetic energy density ${B_2^2 \over 8\pi} \approx 10^{-11} erg/cm^3$
to the kinetic energy density ${1 \over 2} \rho v^2 \approx 10^{-6}
erg/cm^3$. By 300 years the reverse
shock has reached the inner region of the constant density
(Fig.3), and the intershock region has broadened.
As a result, the density at the reverse shock and magnetic field strength
at the contact discontinuity decrease accordingly (see also Fig.4). The
mass ratio at 300 years is about 1.6. 
According to Chevalier (1982), the self-similar solution is valid for
$t < t_c$ where 
\begin{equation}
t_c = 0.36({M^5 \over E_k^3 \rho_a^{2}})^{1/6}
\end{equation}
and $\rho_a$ is the ambient density.  
For our parameters, $t_c = 225 yr $.  Our simulation agrees well 
with this value.
Fig.5 shows the comparison of evolutions between the self-similar
solutions for n=7 and s=0 (Chevalier 1982) and our numerical solutions
at the contact discontinuity.  The numerical solution shows a higher
velocity than predicted by the self-similar solution until 345 years, 
thereafter becoming slower.
As explained by Band and Liang (1988), this slowdown occurs 
because when the reverse shock moves into the inner core of the ejecta
, the density upstream of the reverse shock decreases, and the
force of the ``piston'' weakens.    Note the deceleration of the ejecta is
the strongest at the earliest stage.   This strong deceleration 
directly affects to the evolution of the instability as we will see in
the 2D simulation.

\vspace{5mm}
\subsection{2D results}
\vspace{5mm}

\subsubsection{Evolution of the instabilities}
\vspace{5mm}

  The generalized condition for R-T instabilities can be written as (Jones et
al. 1981)
\begin{equation}
 {\partial ln \rho \over \partial x}{\partial P \over \partial x} < 0
\end{equation}
Basically, this condition states that the local density gradient
${\partial \rho \over \partial x}$, is in the opposite direction of the
local effective gravity $g$, where for local hydrostatic
equilibrium $g = {1 \over \rho}{\partial P \over \partial x}$.
If this condition is satisfied, denser fluid is sitting atop lighter
fluid in an effective gravitational field, and this is unstable.
The region near the contact discontinuity is most unstable (Fig.2) by
this criterion.
At early times, the short wavelengths appear first because the growth
rate is inversely proportional to the square root of the wavelength.
Large fingers become dominant at later stages because larger fingers
feel less drag and reach a higher terminal velocity than smaller
fingers.  The terminal
velocity is proportional to $\sqrt {gr}$ where
$r$ is the radius of the finger.  As the finger
grows, it streams ahead of the shocked ambient gas, and thus the
boundary of the finger becomes sheared.  As a consequence, 
the Kelvin-Helmholtz (hereafter K-H) shear wave instability develops
and mushroom caps, which are regions of high vorticity, are  
generated at the tips of the fingers.  
K-H instabilities on the side of the finger can even detach the finger.
Because of flux freezing,
the magnetic fields are amplified by shearing and stretching at the
finger boundaries (on the other hand, in the case with an initial
toroidal magnetic field, 
interchange modes develop and no amplification of the magnetic fields
occurs because there is no stretching of the field lines.).
Fig.6 shows the grey scale images of density (left) and magnetic
pressure (right) in the shell region at 100,200,300,400,and 500 years 
from bottom to top.   
The wider intershock region is noticeable at 400 and 500 years.  The
wavelength of instability increases as the intershock region becomes
wider.   The thickness of the mixing layer is about
half of the magnetic shell thickness at 500 years where the magnetic
shell is defined as the region between the outer shock and the
innermost boundary of R-T bubbles. 
Our results show well developed mushroom caps at the tips of
fingers due to the K-H instability in good agreement with the purely
hydrodynamic simulations of Chevalier et al. (1992).  
Both the forward shock and reverse
shock are clearly seen and are not significantly influenced by the
instabilities. 
Notice that the reverse shock is not seen in the image of magnetic
pressure because the mixing due to the instability does not reach to
the reverse shock.   

   The effects of compressibility on the evolution
of the fingers is significant.  As fingers approach the forward shock,
they move into a higher pressure environment and are compressed.
For instance, the density at the reverse shock in 1D was about
$4.6 \times 10^{-24} g/cm^{3}$ at 500 years.  In the 2D simulation, a
higher density of $5.5 \times 10^{-24} g/cm^{3}$ is found 
near the tip of some fingers as an evidence of the compression.
As a result, magnetic fields are amplified by compression as well
as stretching.  The strongest magnetic fields are seen along the
sides and tips of the fingers rather than at their base where bubbles
of shocked ambient medium collect. This is because in the fluid frame
the fingers have a higher speed than the bubbles.
A detailed image at 500 years is shown in Fig.7.  It shows, from top
to bottom,  grey
scale images of the density, the magnitude of magnetic field, the 
vorticity, and the current density.  
It is clearly seen that the
vorticity in the wings of mushroom cap is the strongest.  However the
magnetic field is not strong in the strong vorticity region. One might
naively expect that the magnetic field should be strong in that region
as well since
the vorticity and the magnetic field obey the same equations in an
incompressible fluid with a passive field.  This is not the case in
our 2D calculation because magnetic field and vorticity have very
different initial values, the
vorticity has only a component in the perpendicular direction to the
simulation plane,
while there is no B-field in this direction.
The current density is found to be strong where mushrooms collide with
each other.
Current sheets form and magnetic reconnection can occur.
Cattaneo et al.(1990) in a study of bouyancy-driven instabilities, 
also found that magnetic reconnection occurs when
mushrooms detach themselves from the layer and when they collide with
each other.

 The instability is
strongest at about 60 years as we can see in Fig.8.   The turbulent
energy density is defined  as peculiar kinetic energy density
by excluding the contribution from pure expansion of the SNR,
$ E_{tur} = {\int {1 \over 2} \rho (v_{1,p}^2 + v_{2,p}^2) dV \over \int
dV} $ where $v_{1,p}$
and $v_{2,p}$ are defined as $v_1 - \overline{v_1}$ and $v_2 -
\overline{v_2}$, and  $\overline{v_1}$ and $\overline{v_2}$ are the angular
average of $v_1$ and $v_2$, respectively.  The strong
deceleration at the early stage drives the rapid increase of the
turbulent energy.   After the first peak, the turbulent energy density
decreases rapidly due to the weak deceleration, and then it tends to
stay for a while during the self-similar stage (saturation of the
instability).  At around 300 years, the turbulent energy density
starts to decrease again because the remnant evolves beyond the
self-similar stage.  Now the reverse shock enters the constant density
region of an inner core, the intershock region becomes wide, and the
instability weakens.   About 5.8\% of initial kinetic energy
transfered into the 
turbulent energy by the end of simulation.

  The evolution of the magnetic energy
density reflects the evolution of the instabilities directly since the
magnetic field is amplified by the turbulent flows.  The magnetic
energy density is  defined as peculiar magnetic energy density which
exclude the initial magnetic field and shocked magnetic field.  So,
the peculiar magnetic energy density contains the magnetic field
amplified by the instabilities only.
The total magnetic energy density decreases from about
380 years due to the weakened instabilities and the decreased strength
of magnetic fields near the contact discontinuity (see figures 3 and 4).
The evolution of the thickness of a magnetic shell and the thickness of an
intershock region is shown in Fig.9.  
The self-similar stage which can
be noticed from the constant thickness persists until about 220 years.
The big bump of magnetic shell up to
100 years represents the vigorous instability due to the strong
deceleration at early stage.  The image at 100 years in Figure 6
reveals some of bubbles penetrating close to the reverse shock front
which result in a thicker magnetic shell.

\vspace{5mm}

\subsubsection{Strength and Distribution of Magnetic Fields}
\vspace{5mm}

  Initially, the magnetic field is effectively passive as the
supernova energy exceeds the swept up magnetic energy by many orders
of magnitude.   We now ask whether the field becomes dynamically
important as it is amplified.   To answer this question, we compute
the energy spectra.  From Fig.10, we can see that the turbulent
energy density exceeds the magnetic energy density
at all scales.  Therefore, we can say that the magnetic field has not 
reached the equipartition globally on any scale contrary to the usual 
assumption.
In Fig.8 it is also seen that the turbulent energy density is greater 
by about factor of 300 than the magnetic energy density.  The magnetic
energy achieved by the instability is about 0.3\% of the turbulent
energy at the end of simulation.
Even though the field has not reached equipartition globally, it may
be dynamically important locally.  To investigate this, we compare
the turbulent energy density and magnetic energy density
at each cell. We find that the field becomes dynamically important 
at the tips of some fingers and around the fingers although the region
with dynamically important field is only a very small
fraction of the magnetic shell.  This fraction is likely to increase
with the numerical resolution. We discuss this issue further in section 4.

  The angular averages of total magnetic field at 500 years are compared to
the result of 1D simulation in Fig.11.   The 2D result shows the
peak magnetic field strength is near the contact discontinuity due to the
amplification by the R-T instabilities.   We can see 
the generation of radial component of the magnetic fields quite
clearly in the third plot, as well as the amplification of the
tangential magnetic field in the fourth plot.
  In Fig.8, it is found that the radial
component of peculiar magnetic energy density is comparable to
circumferential(tangential) component.   
At 500 years, the radial field becomes the
dominant component with an
energy density of about 1.4 times the circumferential component
(Fig.8) when we consider only the peculiar magnetic energy which is
the result of the instability.
 It seems that the relative magnetic energy
density of the radial component increases as the mixing region becomes
wider allowing the formation of longer fingers.

Fig.12 shows the spectra of peculiar magnetic energy components at
400 years and 500 years. The radial field is dominant in the large scale
structure (mode number, 13 to 18) although the relative strength 
fluctuates depending on the size of the structures.   
It seems that the dominance of radial
field in the large scale in our simulation may be able to explain the net
radial orientation of the magnetic field in young supernova remnants
if the magnetic field in the ambient medium is disordered unlike the
initial condition of our simulation.
It should be noted that the direct quantity to look at
is not the magnetic energy but the radio polarization because the
magnetic energy component is not necessarily proportional to the
polarization.  Since our simulation is restricted to two-dimensional
space, we are unable to address the radio polarization directly.  
A future paper will report on fully 3D MHD simulations of the
Rayleigh-Taylor instability in SNR (Jun 1995; Jun and Norman 1995).
Polarization mapping of our 3D model confirms that peculiar B vectors
in the main radio shell are highly polarized in radial direction. 
At the moment, our 2D simulations only illustrate the main mechanism
responsible for producing radial magnetic fields in young SNR.

\vspace{5mm}
\subsubsection{Radio and X-ray emission}
\vspace{5mm}

  To extract some observational features from our simulations, 
we have computed the radio and X-ray emissivity.  
We assume that X-ray emission comes from
thermal bremsstrahlung while radio emission is due to nonthermal
synchrotron radiation.   For X-ray, we take the total amount of energy
radiated in free-free transitions per $cm^3$ per $second$ integrated
over all frequency (Spitzer 1978)
\begin{equation}
 \epsilon _{ff} = 1.426 \times 10^{-27} Z_i^2 n_e n_i T^{1/2}<g_{ff}>
ergs cm^{-3} s^{-1}
\end{equation}
where $Z_i$ is the atomic number of ion $i$, $n_e$ is the electron
density, $n_i$ is the ionic density, $T$ is the gas temperature, and
$<g_{ff}>$ is the mean Gaunt factor.
The mean atomic number (19.4) of ion in the ejecta is obtained from the
modeling of Tycho's SNR by Hamilton, Sarazin, \& Szymkowiak (1986).
Heavy element abundance is assumed to be constant in the ejecta
material.  The weighted average between the ejecta and ambient
material is computed for the abundance in the mixing region by using
the mass fraction distribution.
In actuality, the detailed X-ray emissivity is more complicated due to the
line emission of heavy elements, nonequilibrium ionization effects,
and so on.  For example, 
it is shown that nonequilibrium effects enhance X-ray emission from
young SNR (Shull 1982).  Our choice of thermal bremsstrahlung for
X-ray emission should be applicable only to energies above 2 keV.

The nonthermal radio emissivity (synchrotron emission) is written as
(Clarke 1988)
\begin{equation}
 i(\nu) \propto \rho^{1 - 2\alpha}p^{2 \alpha}{(B sin \psi)}^{\alpha +
1} \nu^{-\alpha}
\end{equation}
where $p$ is the gas pressure, $B$ is the magnetic field strength, $\psi$ is
the angle between the local $B$ field and the line-of-sight, $\nu$
is the frequency of radiation, and $\alpha$ is the spectral index and
taken as 0.6.    This formula is derived by assuming that the
population density of the relativistic electrons have a power law
spectrum.    Since we are not modeling the acceleration and transport
of relativistic electron, it is assumed that the number of
relativistic electrons is proportional to the gas density and the
electron acceleration efficiency does not change with time.
This assumption requires
that electron diffusion lengths are short enough to
confine relativistic electrons to the same fluid element. One should
note that this assumption could also be violated under the adiabatic
expansion since the relativistic gas has an adiabatic index of 4/3.
We postpone a better modeling of radio emission including the acceleration of
electrons by the shock wave to a future project.   Meanwhile, our
model  implies that the radio emissivity is a strong function of
magnetic field and weak funtion of gas density.
The exact proportionality constant depends on the
fraction of relativistic electrons in the bulk fluid, which is unknown.

 The X-ray and radio emissivities at 500 years are shown
in Fig.13.  The left image is radio emissivity and the
right image is X-ray emissivity.   Generally, the radio
emission is strongest around the R-T fingers because the magnetic field
is most strongly amplified there by stretching and winding.   
On the other hand, the
radio emission is rather faint near the R-T bubbles.
Note that the reverse
shock is not seen in the radio image because of the absence of magnetic
field while it is seen in the X-ray image.   However if the
magnetic fields in ejecta is not negligible as Lou (1994) suggested,
the reverse shock should be seen in radio observations.  The strong
X-ray emission from the R-T fingers are noticeable, which is due to
the high abundance of heavy element in the ejecta material.
Fig.14 shows the surface brightness in radio and X-ray computed by
integrating the emissivity along rays.   In our 2D simulation, we
ignore $\psi$ for the moment.  The realistic emission map by
considering viewing angle and local $\psi$ can be produced in 3D
simulation (Jun and Norman 1995).  Each plot corresponds
to the surface brightness at 100, 200, 300, 400, and 500 years from
top to bottom.  The radius is normalized by the radius of the outer
shock and brightness is normalized by the maximum brightness.
The bright shell exists inside of the outer shock in both radio and
X-ray.  The radio emission shows more fluctuation in the structure than X-ray.
The surface brightness at the outer shock is very weak due to the limb
darkening effect.  Radio surface brightness
peaks near the contact interface while X-ray surface brightness peaks near the
reverse shock but they are close to each other.   As the remnant
ages, the peak X-ray surface brightness near the reverse
shock decreases relative to the brightness at the forward shock 
as the density decreases.   The main X-ray shell becomes broader as
the remnants gets older due to the broader mixing layer in the
non-self-similar stage.    Therefore, the thickness of the X-ray shell
is an indicator for the remnant's age.

  Fig.15(a) shows the evolution of radio luminosity.
Radio luminosity is computed by integrating radio emissivity (see
equation 8) over the
volume considering $\phi$-symmetry, that is $ L_{radio} = 
\int \int \int i(\nu)r^2 sin
\theta dr d\theta d\phi = 2\pi \int \int i(\nu) r^2 sin \theta dr
d\theta $.    Although this quantity is not
exactly the true radio luminosity because we ignored the $\psi$
dependence in the emissivity, it is
still closer to the observed radio luminosity than the integration of
emissivity over the computational plane.
The dotted line represents the luminosity from the main shell where the
instability occurs. This is chosen to be the region up to 
$r_{shell} = 0.91 r_{shock}$ where the R-T fingers reach the farthest.
The dashed lines represent the
luminosity from the region between $r = r_{shell}$ and $r = r_{shock}$.
The total radio luminosity
increases until about 375 years ($\mu = 2.4$) and then it starts to
decrease.  The radio luminosity from the instability shell region
(main shell)
always dominates over the radio luminosity from the forward shock
region.  The
decrease of radio luminosity from the main shell at late times is attributed
to the weakened magnetic field strength (1D effect) near the contact 
discontinuity (see Fig.3) and the weakened instability (2D effect) 
after the remnant leaves the self-similar stage.  The radio luminosity
from the forward shock region stops increasing at about 300 years due
to the decreased pressure behind the forward shock (cf. equation(8)).
The evolution of X-ray luminosity which is only applicable to energies
above 2keV is shown in Fig.15(b). X-ray
luminosity is also computed by integrating X-ray emissivity over the
volume, that is $ L_{X-ray} = 2\pi\int \int \epsilon_{ff} r^2 sin 
\theta dr d\theta$.
A peak of X-ray luminosity appears earlier than that of radio
luminosity because the X-ray emission is sensitive to the density
decrease near the reverse shock after the self-similar stage.
The X-ray luminosity is dominated by the main shell due to the
heavy elements in the ejecta material.
Our model predicts that the time history of
each luminosity component is another indicator of the remnant's
dynamical age.
The decline of the radio luminosity at late times is also seen by Dickel et
al. (1993) in their modeling of clumpy circumstellar medium.  They
explain that the decline is due to the result of the different shock
expansion rates which result from different encounters by the shock
with shell clouds.  In our uniform ambient medium R-T model, it is
due to the decreased mixing which occurs beyond the self-similar stage.

  Because we are concerned about our ability to reliably model the
physics of the mixing layer (see discussion on ``numerical
limitation'') which, as we have shown, dominates the radio and X-ray
emission at early times, we ask what fraction of the radio emission
comes from the strongest amplified fields.
  Fig.16 shows the radio luminosity distribution as a function of
magnetic field strength at t=400 years.  
The top histogram shows the luminosity distribution emitted from the whole
remnant and the bottom one shows only the contribution from the
peculiar magnetic fields which is computed by subtracting magnetic
fields from the angular averaged magnetic fields.  These peculiar
components of magnetic fields results from the instability.
About 55 \% of 
total luminosity comes from the range between $B=9.2\times 10^{-6} G$ 
and $B=1.45\times 10^{-5} G$ in the simulation with resolution
300x400.  The maximum
strength of magnetic field is $2.1\times 10^{-4} G$ at this time.  
Therefore the bulk of the radio luminosity comes from relatively
weak fields. 
The contribution of peculiar components of magnetic fields to the
radio luminosity is about 42 \%.   On the other hand, about 58\% of the
total luminosity is within the range between $B=9.2\times 10^{-6} G$ 
and $B=1.45\times 10^{-5} G$ in the simulation with resolution
180x200. And the contribution from peculiar magnetic fields is about 38\% of
the total luminosity in this resolution.  This result shows a tendency
in which the higher resolution simulation produces a greater proportion of 
the radio luminosity due to the instability than the lower resolution
simulation. 
The bottom histograms show the luminosity distributions from the
peculiar magnetic fields for two different resolutions.
In general, the magnetic fields due to the instability (peculiar
component) is stronger than the simple shocked fields.   
Increasing the resolution tends to move the luminosity distribution to
the stronger magnetic fields.  Whether strong magnetic fields (here a strong
magnetic field means the field stronger than the shocked magnetic field.)
will dominate over total luminosity as the resolution increases
further is unclear.  It will depend on not only the efficiency of
magnetic field amplification but also the volume of the mixing layer over
the whole radio shell.

\vspace{5mm}

\section{Numerical Limitations}
\vspace{5mm}

 Our goal is to construct realistic models of Type-I SNR.   In this
work we have explored for the first time multidimensional nonlinear
effects which strongly influence the radio and X-ray properties of
young SNR.   Nontheless, our current
simulation has several numerical limitations which require further
improvement for the better understanding of SNR, which we now discuss.
First, our simulation is restricted to two dimensional space.   As
studied by JNS, 3D simulation of the classical R-T instability
generally produces a somewhat broader mixing
layer (about 40 \% broader than 2D) due to greater finger penetration
in 3D than in 2D.   If the R-T bubbles
reach the reverse shock in 3D, they may distort its shape.
Generally, mixing in 3D simulations is quite different due to the
additional degree of freedom (see JNS).
We also expect that 3D simulations may produce a more dominant radial
field than 2D.
In addition, 3D simulations may produce stronger
magnetic fields just as we found in the study of the classical MHD R-T
instability (JNS).   
Stronger magnetic fields in 3D will result in a higher surface 
brightness as an observational consequence.   3D simulations analogous
to the ones presented here have been recently carried out by us, 
and we will report on them in a subsequent paper. 
Second, we have not included any physical resistivity since we solved
the ideal MHD equations.   The magnetic Reynolds number is a function
of the magnetic
diffusivity and its magnitude influences the amplification of
magnetic fields.  In addition, the magnetic Reynolds number determines the
speed of magnetic reconnection.
In our ideal calculation, the grid resolution sets a numerical
magnetic Reynolds number which increases as the inverse of the cell
size to some power $\geq 1$, which in turn is set by the order of
accuracy of the algorithm.

To diagnose the numerical difficulty in studying the turbulent
amplification of magnetic fields in real SNR, we estimate the 
numerical magnetic Reynolds number in our
simulation by using the relation between the magnetic fluctuations and
the magnetic Reynolds number (Vainstein \& Rosner 1991) :
\begin{equation}
 <B^2> \approx  R_M <\vec B>^2 \nonumber
\end{equation}
where $R_M$ is the magnetic Reynolds number.  This relation is valid
for passive magnetic fields, which is close to our case.  By comparing
the relative magnitudes of $<B^2>$ and $<\vec B>^2$ in the mixing
layer of our simulation, we derive $R_M \approx 50$.  This number is
many orders of magnitude smaller than in nature, yet large enough to
permit the amplification of magnetic fields.
The magnetic Reynolds number $R_M$ is
defined as ${vl \over \eta}$ where $v, l$, and $\eta$ are a typical
speed, a typical length scale, and magnetic diffusivity, respectively.
The Spitzer classical magnetic diffusivity $\eta$ is given by $10^{13}
T^{-3/2} cm^2 s^{-1}$.   Using typical numbers for the SNR case, $T
\approx 10^7 \sim 10^8 K$, $v \approx 10^8 cm/s$, and $l \approx 0.1 pc $, we
obtain a 
magnetic Reynolds number of about $10^{24}$.  This number is much larger
than the typical solar magnetic Reynolds number $10^{7}$, and vastly
larger than what can be simulated directly.    This result is not as
discouraging as it seems, for two reasons.   First, we have estimated
$R_M$ using the classical value for $\eta$, and this ignores anomolous
diffusivities which are orders of magnitude larger in diffuse plasmas
because, for a turbulent plasma, the collision time and corresponding
electrical conductivity can often be much smaller than the Spitzer
values (Priest 1982).   Second, while although we are not treating the
small scale magnetic fields correctly, we believe we are modeling the
large scale fields correctly.  This means that the radio luminosity which is
sensitive to the small scale field, is underestimated, whereas the
radio polarization which is determined by the large-scale field is
accurately computed.  Insofar as the radio luminosity is uncertain to
within the normalizing factor in eq.(8), whereas we are primarily
interested in understanding the polarization properties of young SNR,
our simulations are still providing us some useful new results.
We expect that the volume occupied by the dynamically important fields
may increase in higher resolution 3D simulations.
However, it is still uncertain if equipartition between
turbulent energy and magnetic energy is obtained on the
time scale of a real SNR.  The equipartition may be achieved at least in small
scales since the eddy-turnover time is shorter for small scale structures.
Despite all of numerical difficulties, we believe that our current
calculation still serves a good basis for understanding global features
such as {\it large scale structures of magnetic fields} which was one of our
original motivation for this study.

\section{Discussions and Conclusions}
\vspace{5mm}

  We defer a detailed comparison with observations to our forthcoming
paper presenting the results of full 3D simulations of young SNR (Jun
\& Norman 1995).  However, some useful comparisons can already be made on
the basis of the 2D simulations.
In Fig.11, the angle-averaged
magnetic fields shows a maximum value about $2 \times 10^{-5} G$
near the contact discontinuity.  This is about one order of magnitude
lower than the estimated value $3 \times 10^{-4} G$ in Tycho's SNR
(Strom and Duin 1973).  As we discussed in Sec.4, this is very likely
an effect of numerical resolution limiting the amplification of
magnetic fields on small scales.
In addition, it should be noted that the estimated
strength of magnetic field in young SNR is only a rough value because
the minimum energy requirement is used to obtain it.  The condition
for minimum energy requirements corresponds closely to the condition
of equipartition of energies in the relativistic particles and the
magnetic field (e.g. see Longair 1992).  However, there is no physical
justification for the equipartition.  The information about a 
filling factor (the fraction of the volume occupied by radio
emitting material) is an another uncertainty in the observation.
 Since our result of magnetic-field strength is much lower than the
strength estimated by the equipartition assumption in young SNRs, the
higher energy of relativistic electrons is required to account for the
observed radio luminosity of young SNRs.

  The surface brightness predicted from our simulation shows that main
shells of 
radio and X-ray emission are close to each other but not exactly coincident.
This result may explain the coincidence of the inner radio shell 
and X-ray emission in the main shell of Tycho's SNR.    However neither
our simulation nor observation show detailed correspondence between
radio and X-ray features.  This is explained in our model as a simple
result of the nature of the strongly emitting regions in these two
wavebands : the X-rays are emitted by the dense fingers of stellar
ejecta, whereas the radio emission preferentially samples the
strong magnetic field lines draped around the R-T fingers.

 One of the most puzzling features in young SNRs is the bright rims
in Tycho's SNR and the SNR of A.D. 1006 which our model does not explain.
Dorfi (1990) has studied the cosmic-ray driven instability in young
SNR, and has found that the region in the precursors of two
shocks is unstable.  However, this instability generally requires a dominant
cosmic-ray pressure which is in doubt (Markiewicz et al. 1990; Drury
et al. 1995). One can also ask why rims are not seen in all young
SNRs if this instability is a generic feature of the shock in SNR.
The detailed hydrodynamical simulation of this instability 
in multi-dimensions should be pursued.  
Another possible mechanism is the clumpy medium model(e.g.
Dickel et al. 1989). Although we don't know have detailed information about the
inhomogeneity in the ambient medium, it is worth studying the
effect of the clumpy medium in general.   This model has been
investigated very recently with encouraging results (Jun 1995).  The
clumpy medium model is found to produce a thicker and clumpier mixing
layer than a uniform medium model with a dominant radial component of
magnetic fields.

  We have carried out the first 2D nonlinear MHD simulations of a
young supernova remnant propagating into a uniform density, uniformly
magnetized medium.  Our moving grid technique has allowed us to study
the instability of the mixing layer between the stellar ejecta and the
swept up ISM in both the Chevalier (2-shock self-similar) and
non-self-similar regimes.  Summarizing our main results, we find :

1. The R-T and K-H instabilities in the mixing layer amplify the
existing magnetic fields locally by as much as a factor of 60 around
dense fingers by the stretching and the compression of active
fingers.  Globally, the angle-averaged magnetic fields near the
contact discontinuity increased to about $2 \times 10^{-5}G$ from an
ambient value of $3.5 \times 10^{-6}G$.

2. Equipartition between the turbulent energy density and magnetic
energy density is not achieved globally.   The magnetic energy density
reaches only about 0.3 \% of the turbulent energy density by the end
of simulation.  The amplified field becomes
dynamically important locally around the active finger.  
However the magnitude of amplification is sensitive to numerical resolution.

3. The R-T instability model produces thick clumpy radio and X-ray
(thermal bremsstrahlung) shells inside of the outer shock.  
The X-ray emission peaks inside of
the radio emission but they are close each other.   
Our model shows that there is no detailed correspondence
between substructure in the radio and X-ray emission. 

4. The time history of luminosity reflects the dynamics of the
remnant and the instability.
Both the radio and X-ray (thermal bremsstrahlung) luminosities
increase during the self-similar stage but then decrease as the remnant
evolves beyond the self-similar stage.

5. Our simulation produces comparable or dominant (at late stages) 
radial magnetic fields in the main radio shell.  However, our model
cannot account for radial magnetic fields adjacent to the outer shock front,
which may require additional instabilities to generate.

\vspace{5mm}

   We are grateful to Fausto Cattaneo, John Dickel, Tom Jones, 
and Bob Rosner for the useful discussion.   
We also thank Roger Chevalier for comments.  
A number of constructive comments by the referee, Steve Reynolds were
particularly useful.
The simulations were done on the Cray C90 at the Pittsburgh
Supercomputing Center.

\bibliographystyle{plain}

\vfil\eject

\section{Figure Captions}

%\begin{figure}
%\centerline{\psfig{figure=fig2.eps,width=5in}}
%\caption{\tenrm Linear growthrate of Rayleigh-Taylor instability. 
%The density of a heavy fluid is 20 and the density of a light fluid is 1. 
%The gravitational constant is taken as 1.}
%\label{fig:growthrate}
%\end{figure}

Figure 1.  A schematic representation of a typical young SNR in our
computational space.

Figure 2.  One dimensional result of the evolution of SNR at 100 yrs.
log(density) is the logarithm of gas density, v1 is the radial
velocity, B2 is the tangential component of magnetic field, and p is
the gas pressure. $\mu = 0.23$.

Figure 3.  One dimensional result of the evolution of SNR at 300
yrs. $\mu = 1.6$.

Figure 4.  One dimensional result of the evolution of SNR at 500 yrs.
$\mu = 3.94$.

Figure 5.  The comparison of deceleration constant between our
numerical simulation and Chevalier's self-similar solution.  The
velocity is taken at the contact discontinuity. log(g) is the
logarithm of deceleration constant and log(velocity) is the logarithm
of the velocity at the contact discontinuity.

Figure 6.  The grey scale images of the gas density (left images) and magnetic
pressure (right images) at 500,400,300,200,100 years from top to
bottom.  The black color corresponds to the maximum value in the data.

Figure 7.  The grey scale image of the gas density, the magnitude of
magnetic field, the vorticity, and the current density in the shell region 
at 500 years from top to bottom.  The black color corresponds to the 
maximum value in the data.  In the vorticity and current density
images , white (black) represent negative (positive) values, respectively.

Figure 8.  The evolution of a peculiar magnetic energy density and a turbulent
energy density.  Emag1 is the radial
component of the peculiar magnetic energy density, Emag2 is the circumferential
component of the peculiar magnetic energy, Emag is the total peculiar
magnetic energy 
density, Etur1 is the radial component of the turbulent energy
density, Etur2 is the circumferential component of the turbulent
energy density, and Etur is the total turbulent energy density.  

Figure 9.  The evolution of the thicknesses of a magnetic shell and
an intershock region.   The thickness is normalized by the radius of the
forward shock front. 

Figure 10.  Energy spectra at 400 years.  log(energy) is the logarithm of
each energy at the region of the instability.  In legend, kinetic energy is the
turbulent energy density while magnetic energy is the peculiar
magnetic energy density.

Figure 11. Angular average of the magnitude of the magnetic field at 500
years.  The first plot is the tangential component of a magnetic field
from a 1D simulation, the second plot is the angular average of the
magnitude of a magnetic field from the 2D simulation, 
the third plot is the angular average
of the magnitude of radial component of a magnetic field, and the
fourth plot shows the angular average of the magnitude of tangential
component of a magnetic field.

Figure 12. The spectra of peculiar magnetic energy component at 400 years(top
graph) and 500 years(bottom graph).

Figure 13. Grey scale images of radio (left) and X-ray (right) emission
at 500 yrs.

Figure 14. Surface brightness profiles of radio and X-ray.    The
radius is normalized by the radius of a forward shock.  The units of
surface brighness are arbitrary.  Each plot corresponds to the surface
brightness at 100, 200, 300, 400, and 500 years from top to bottom.

Figure 15. (a) Time history of radio luminosity. (b) Time history of
X-ray luminosity. In legend, subscripts shell
and shock represent the contribution from the main shell of the instability
region and the contribution from the forward shock region, respectively.

Figure 16. The histogram of radio luminosity distribution at t=400
years. Top histogram is the luminosity distribution from entire
region while bottom histogram shows the luminosity from peculiar
magnetic fields. The peculiar magnetic fields are obtained by
subtracting magnetic fields from angular averaged magnetic fields.
The vertical axis represents the partial luminosity divided by the
total radio luminosity.

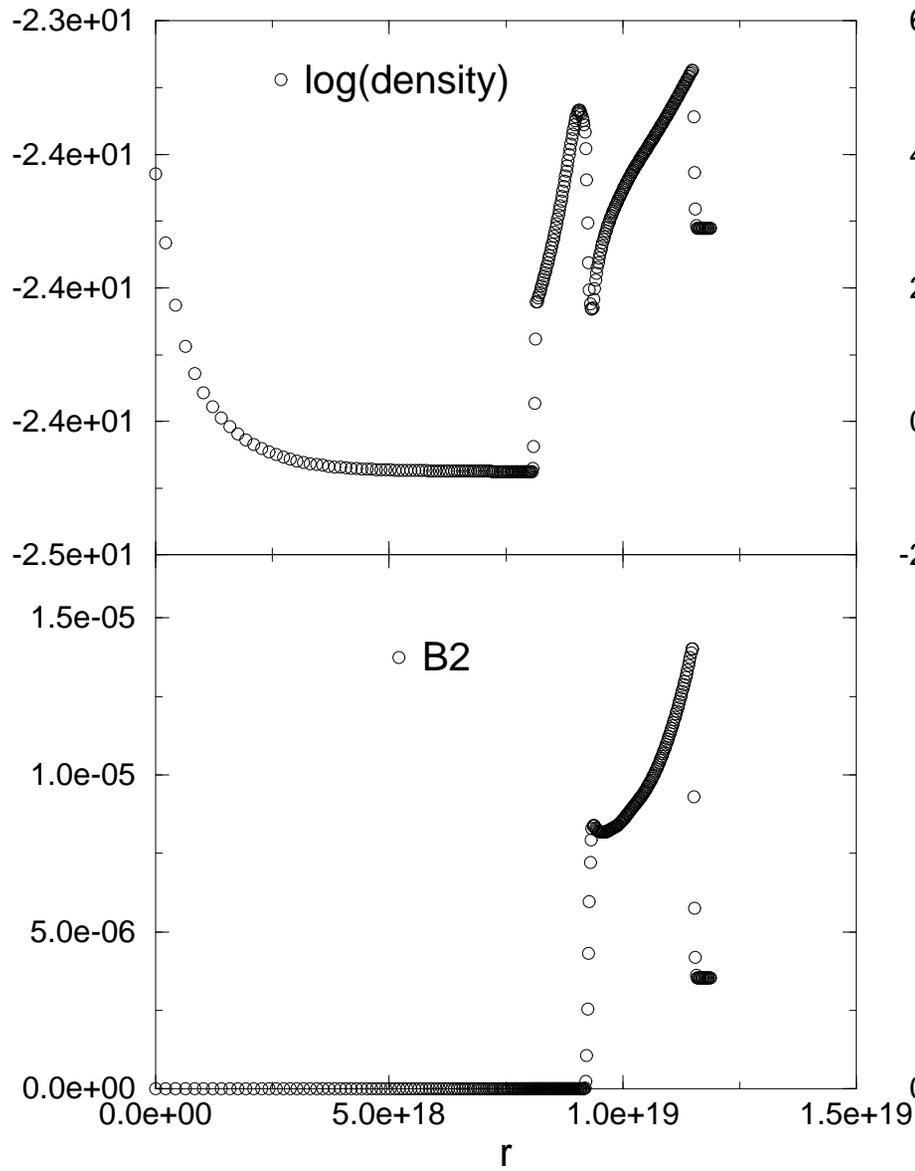
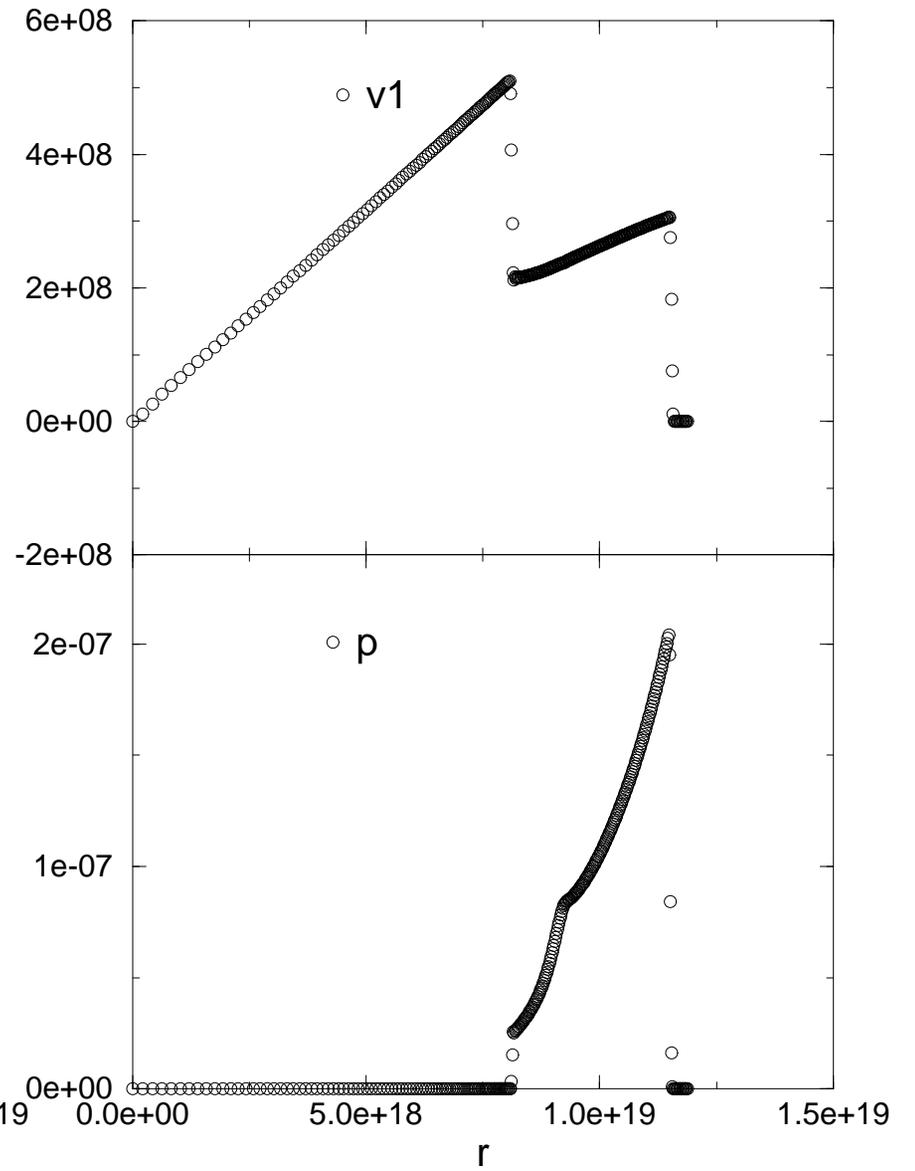
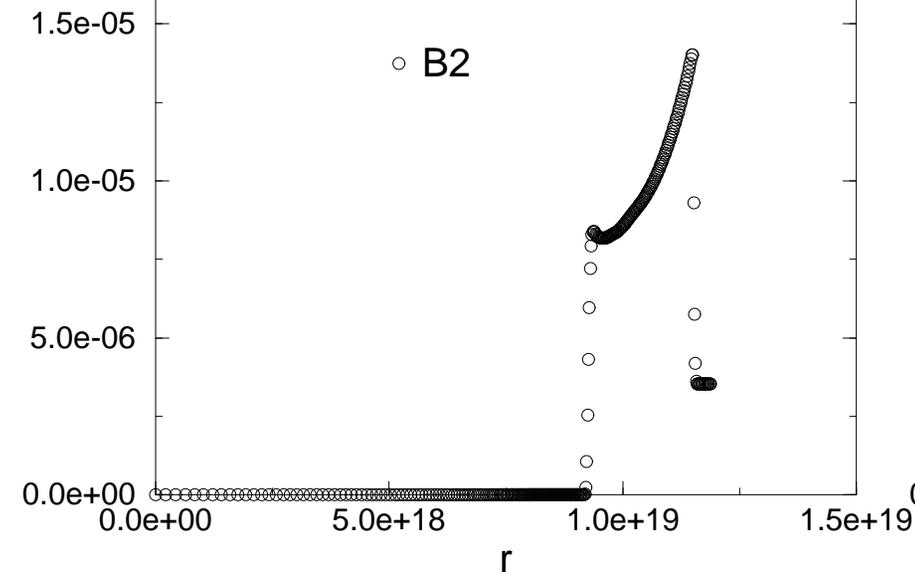
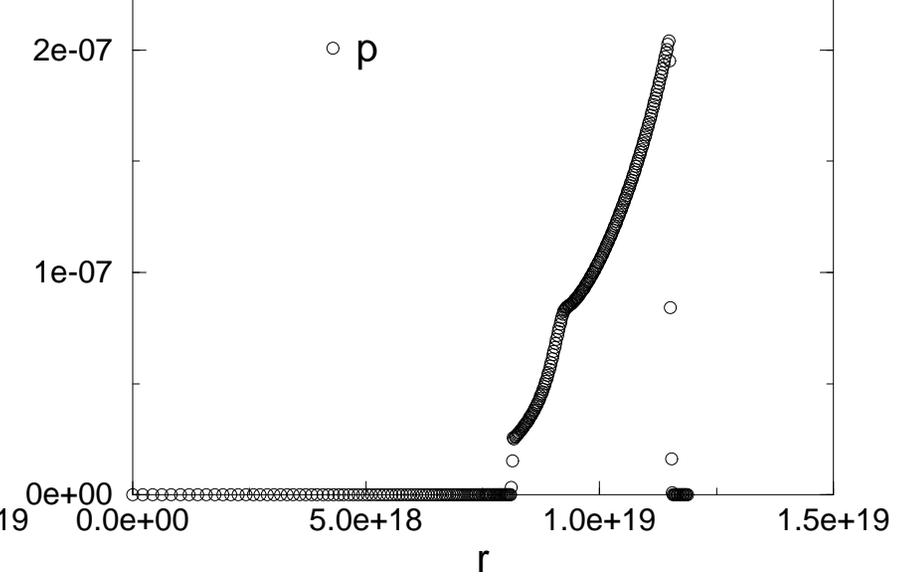
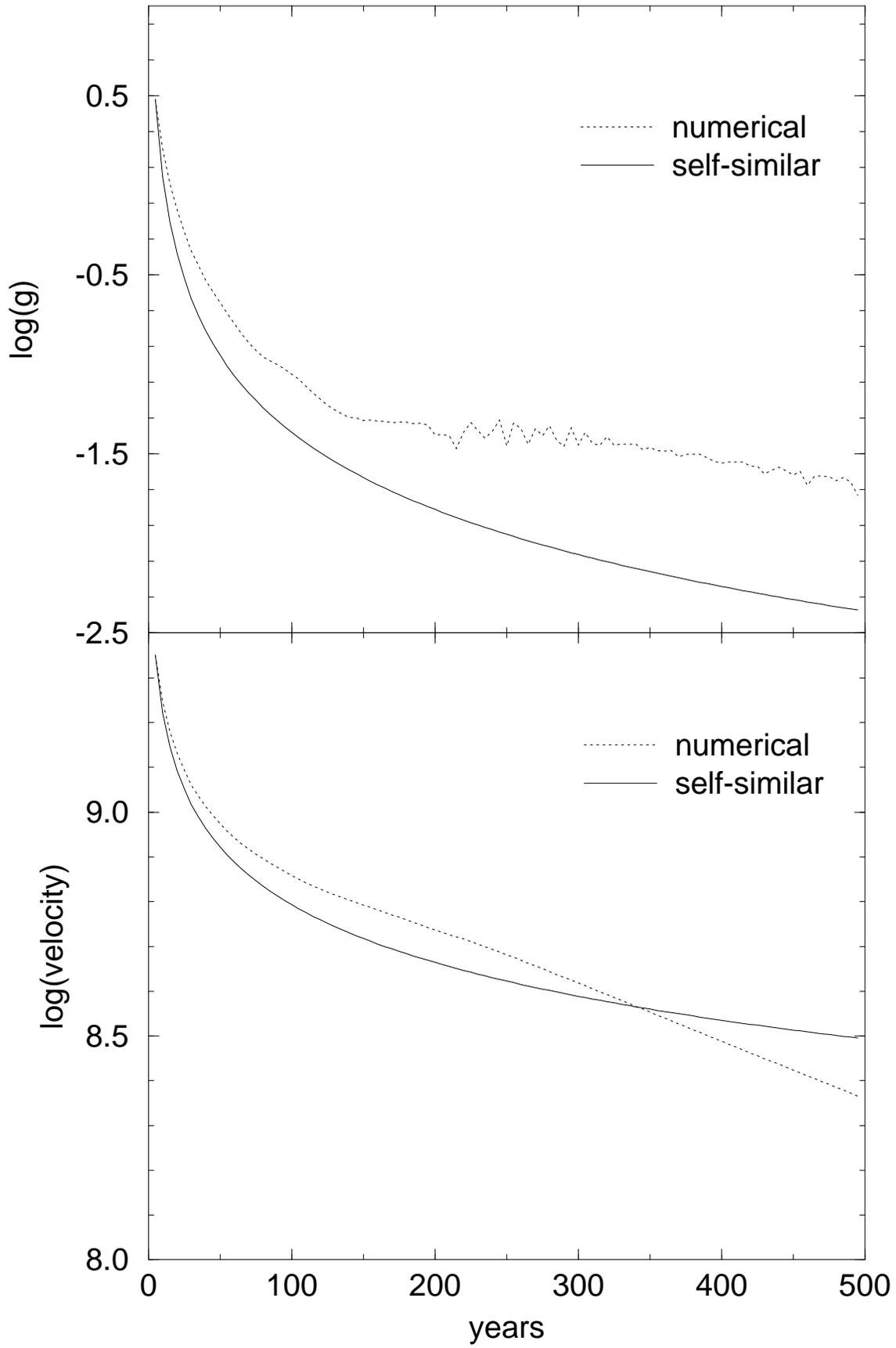
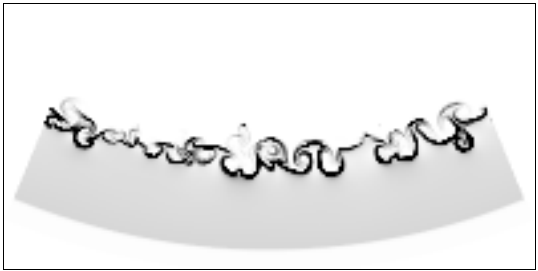
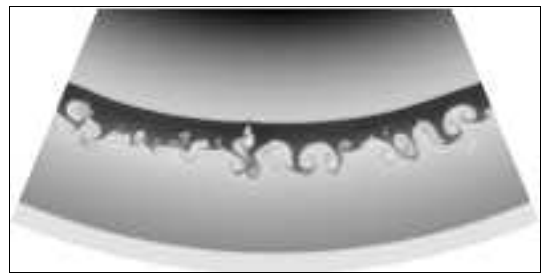
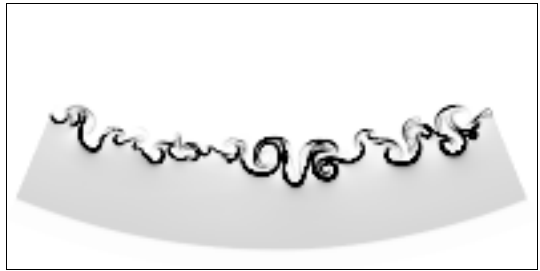
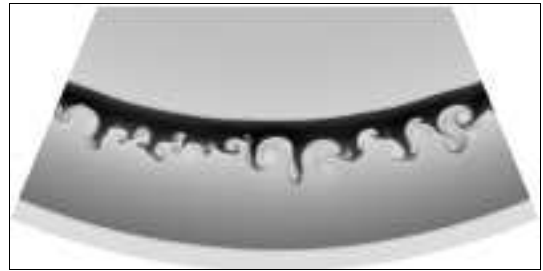
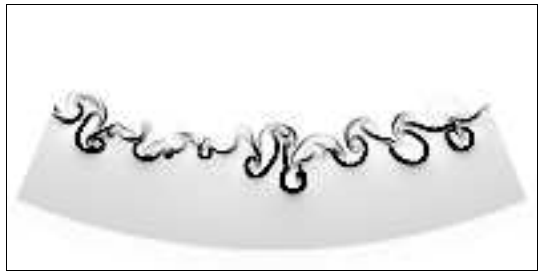
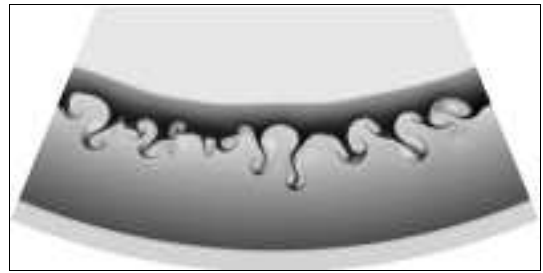
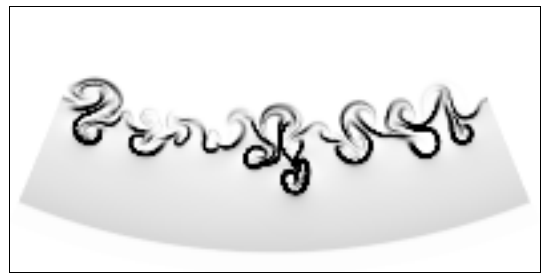
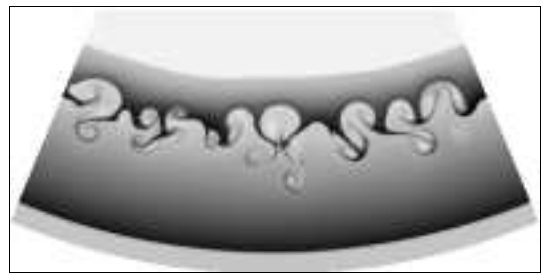
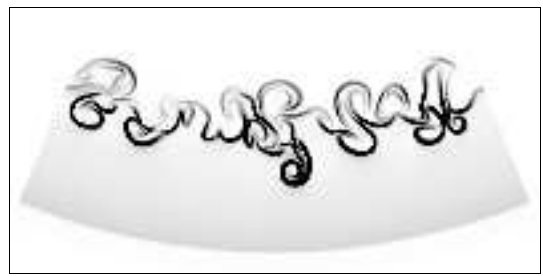
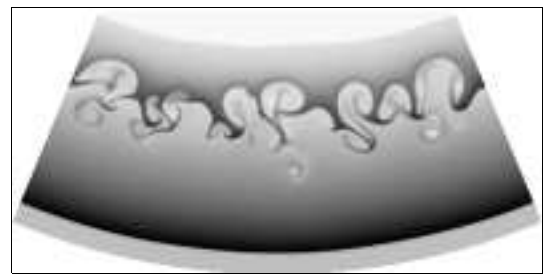
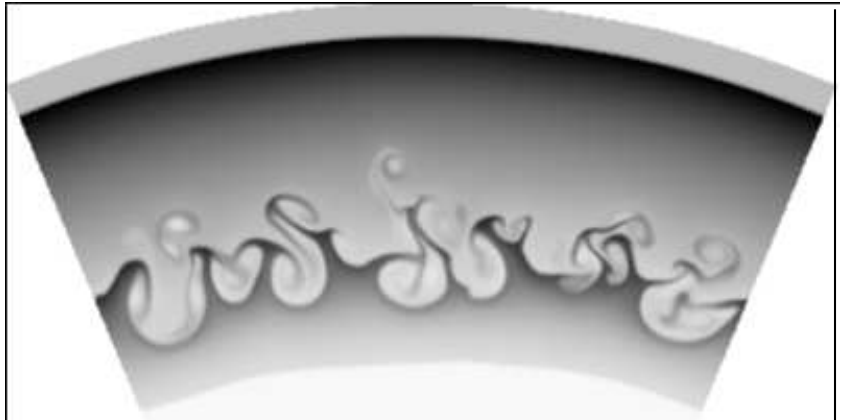
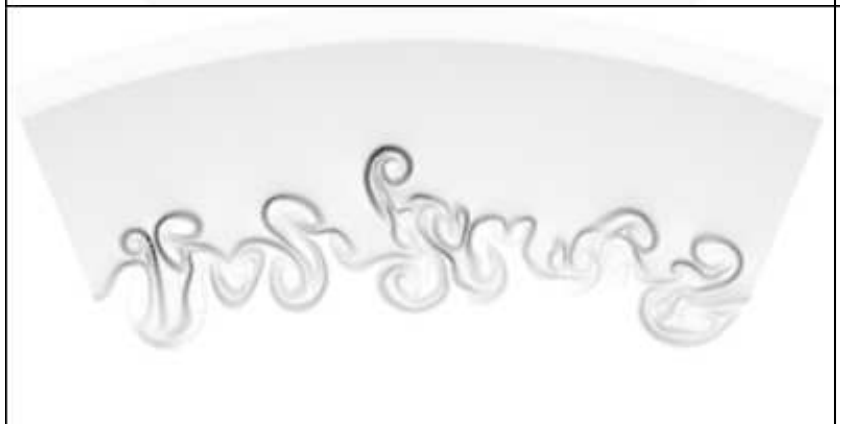
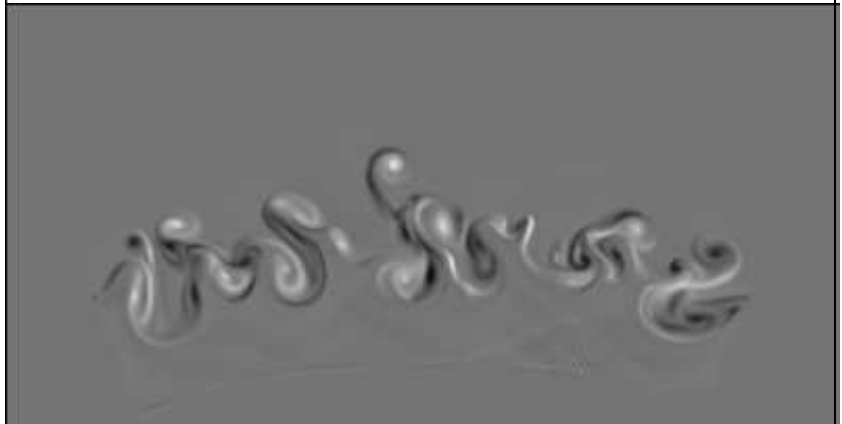
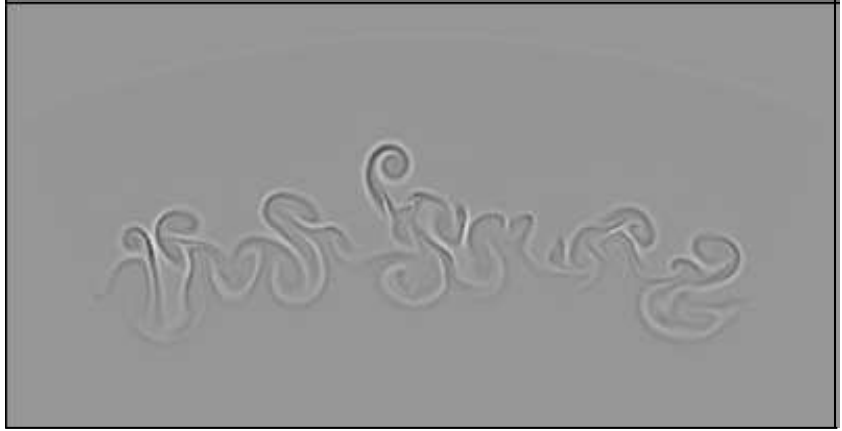
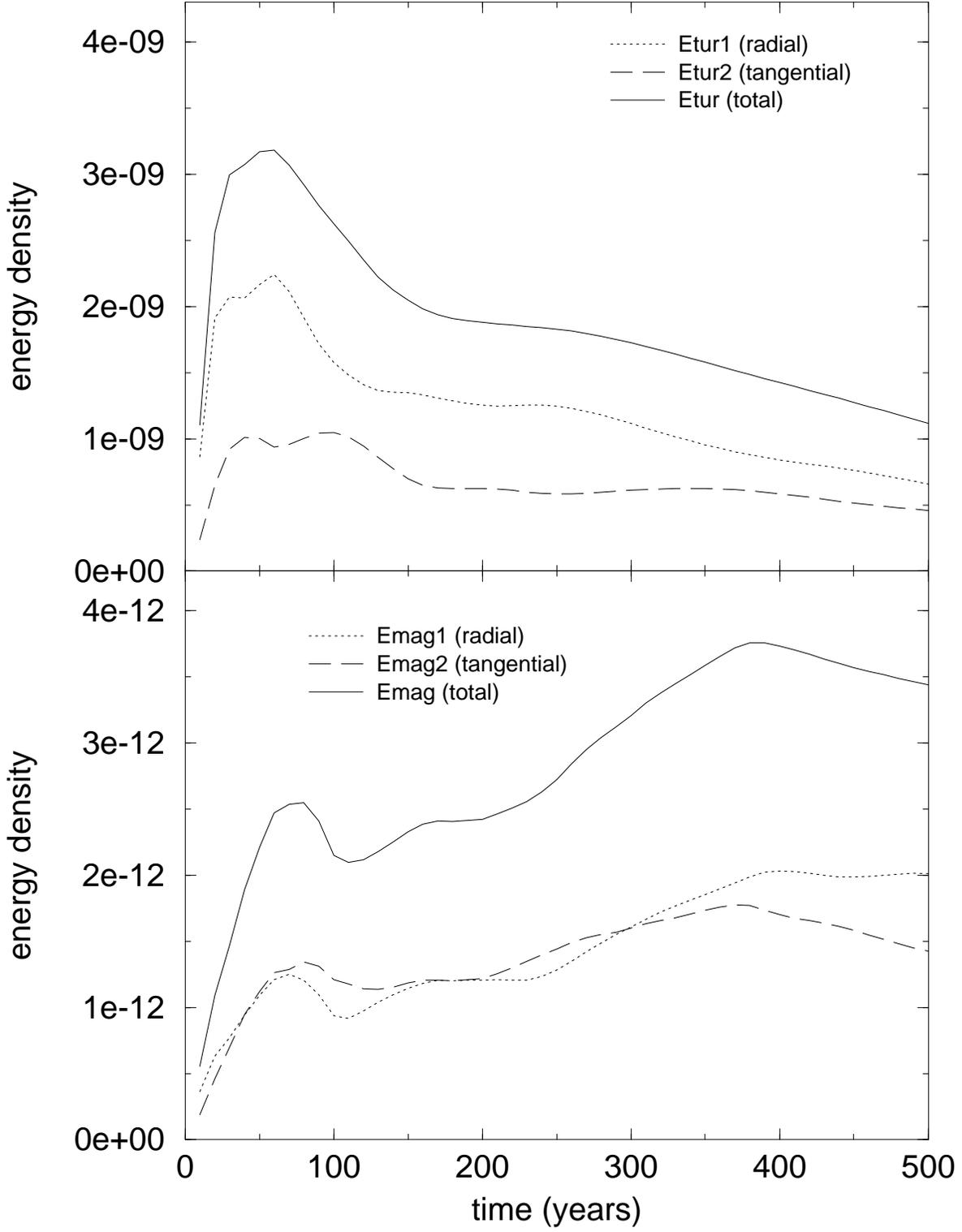
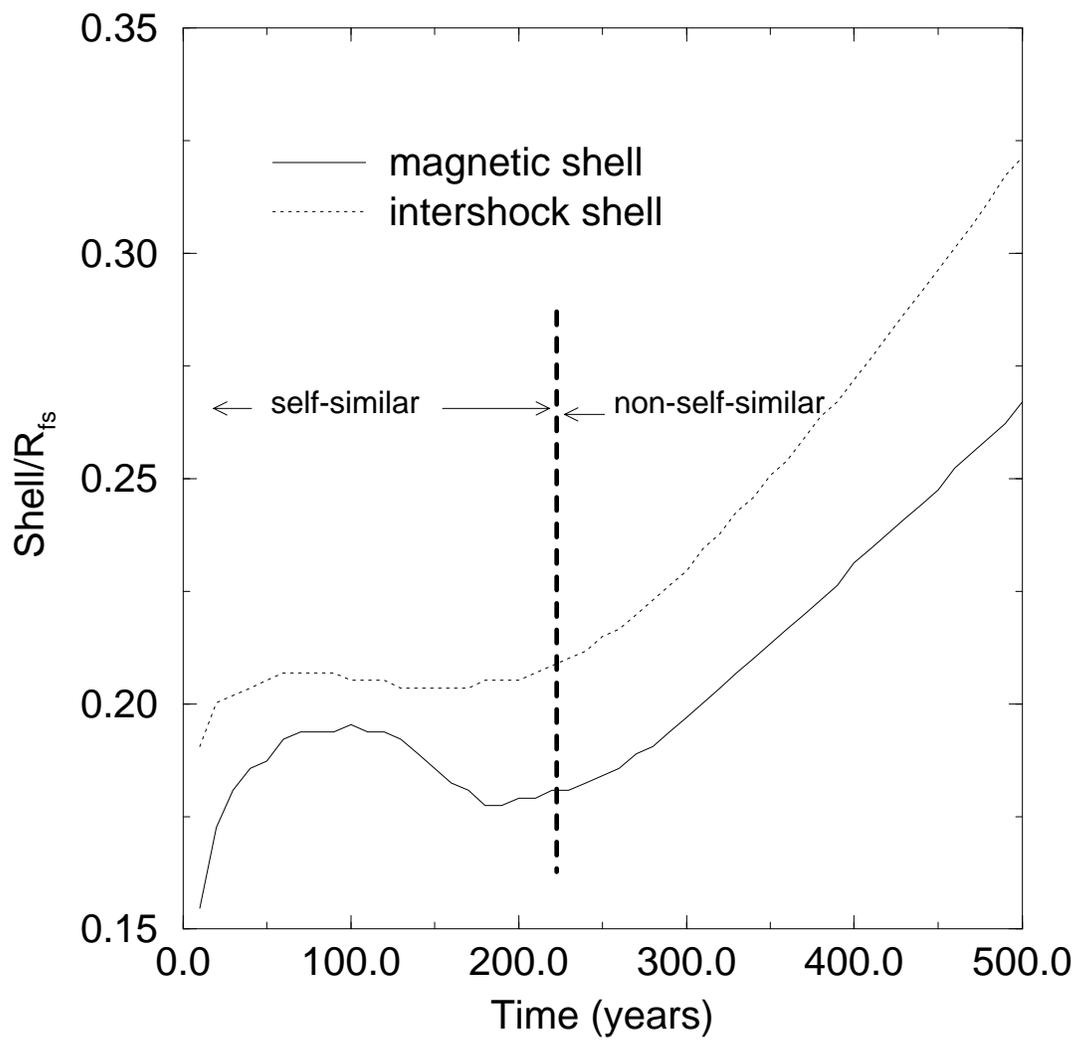
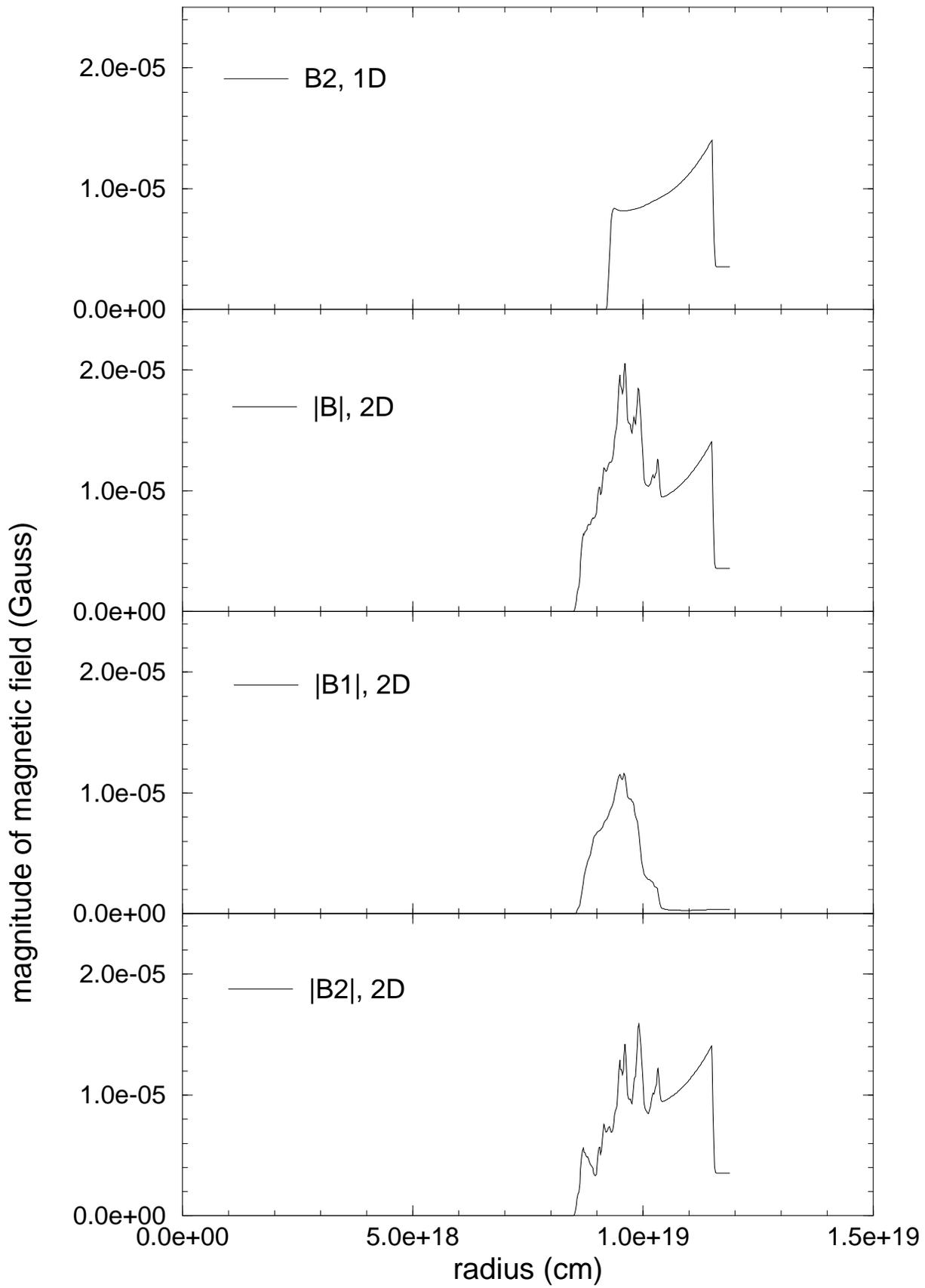
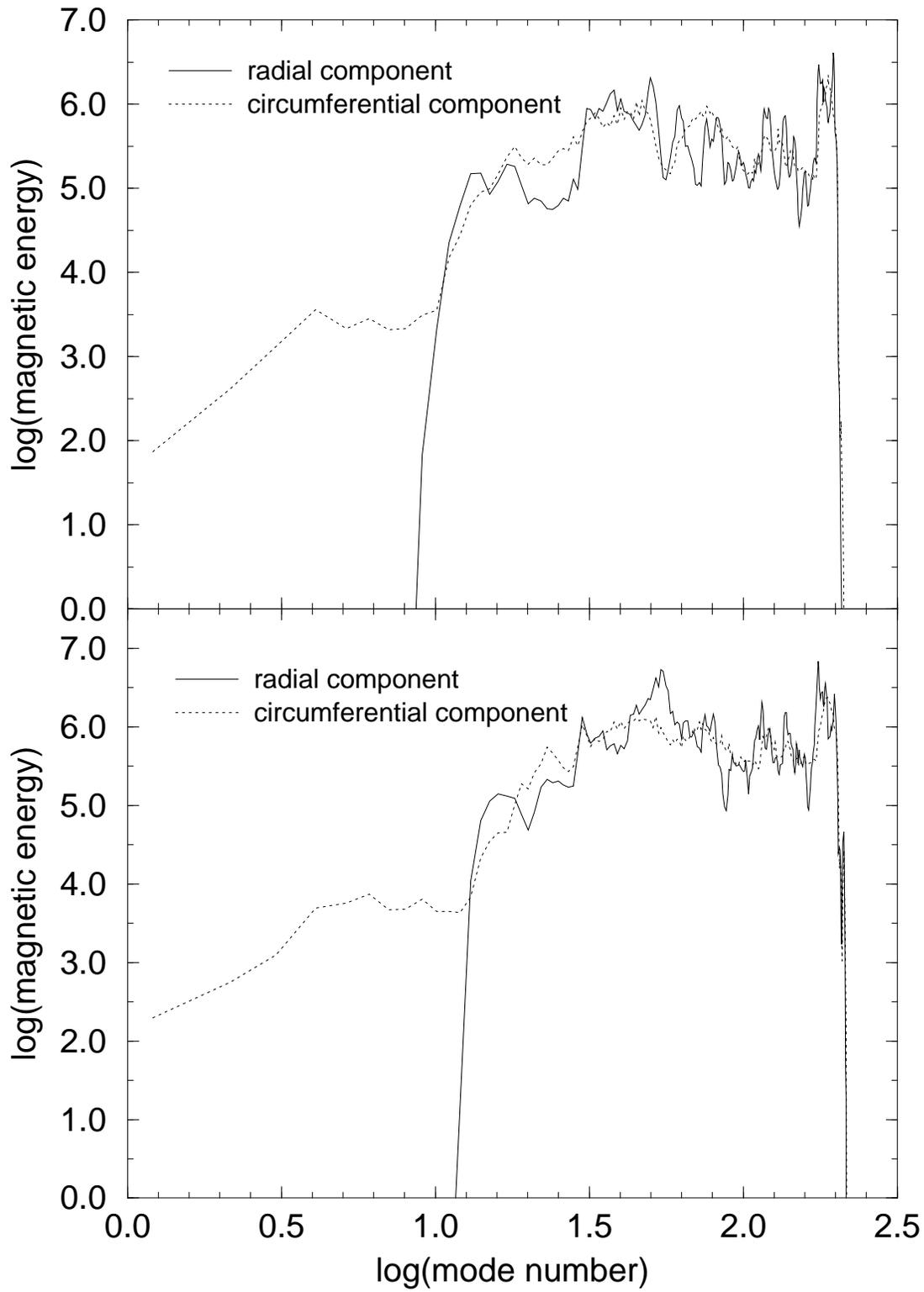
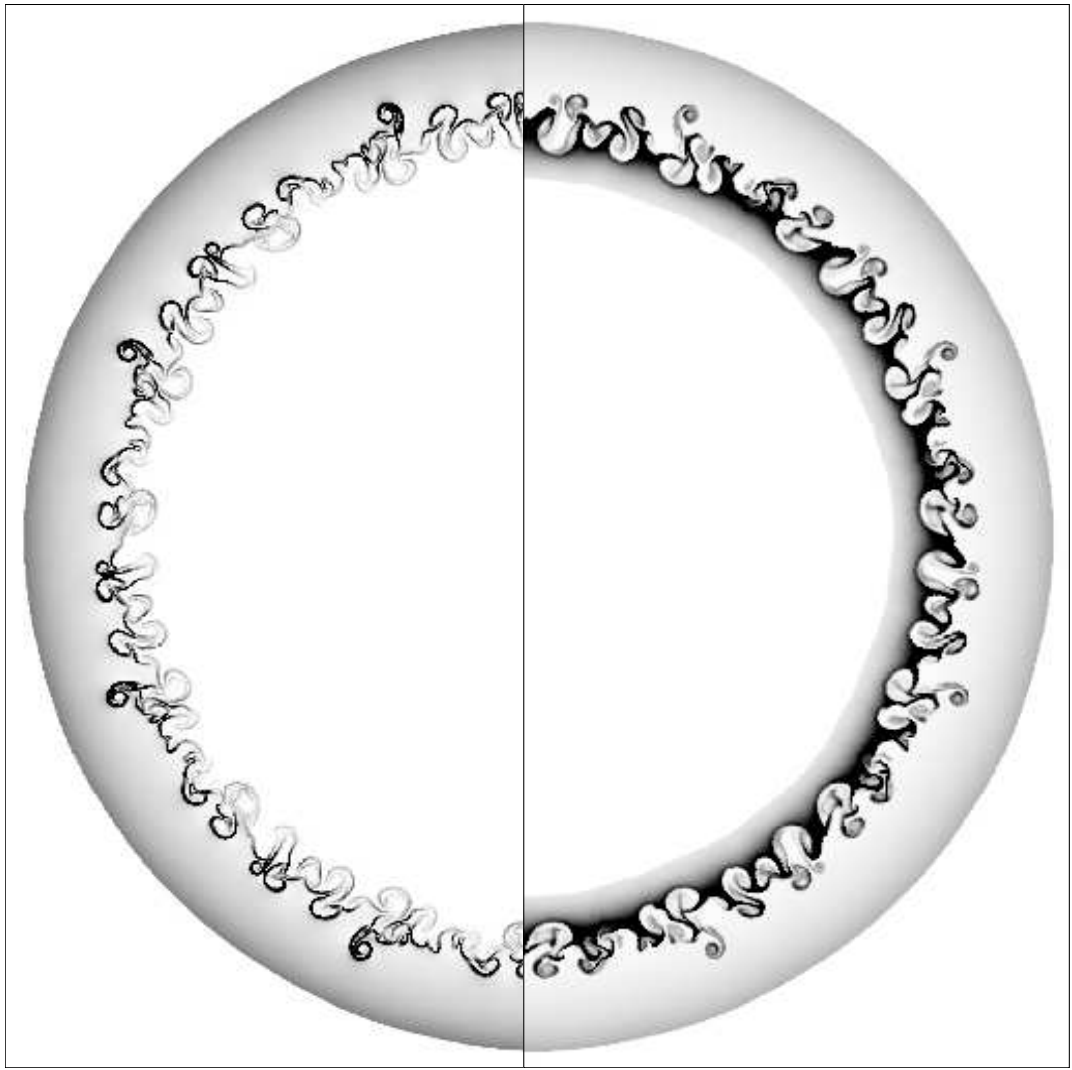
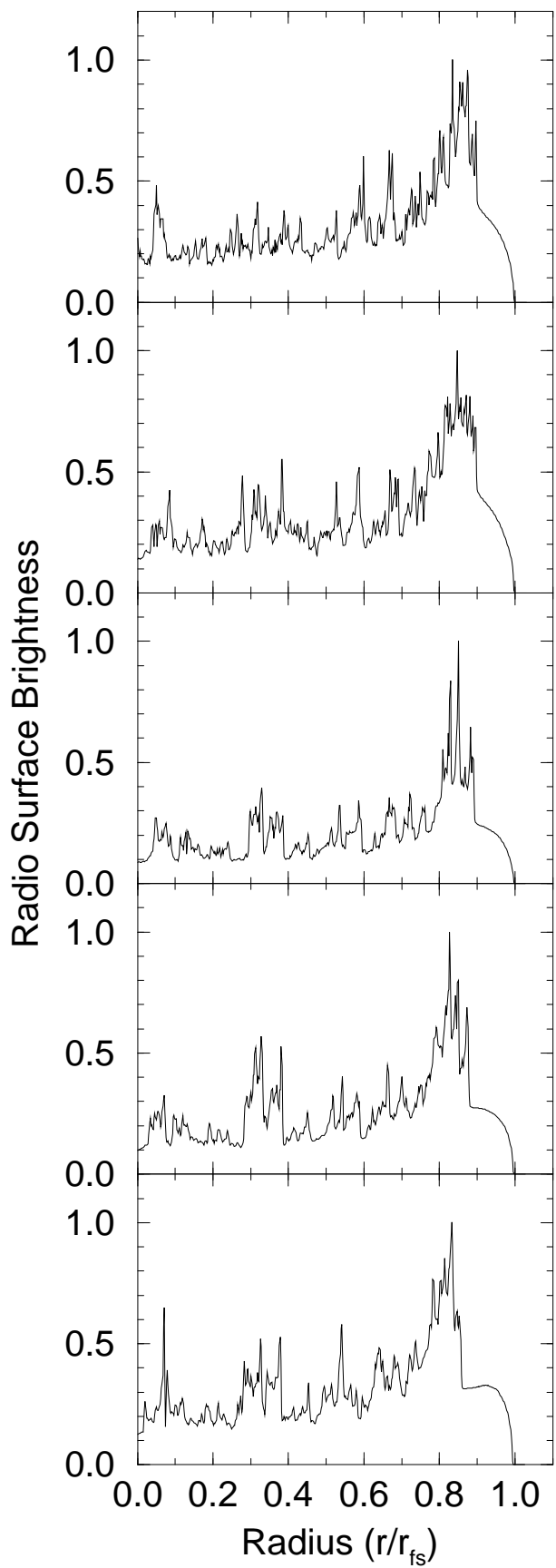
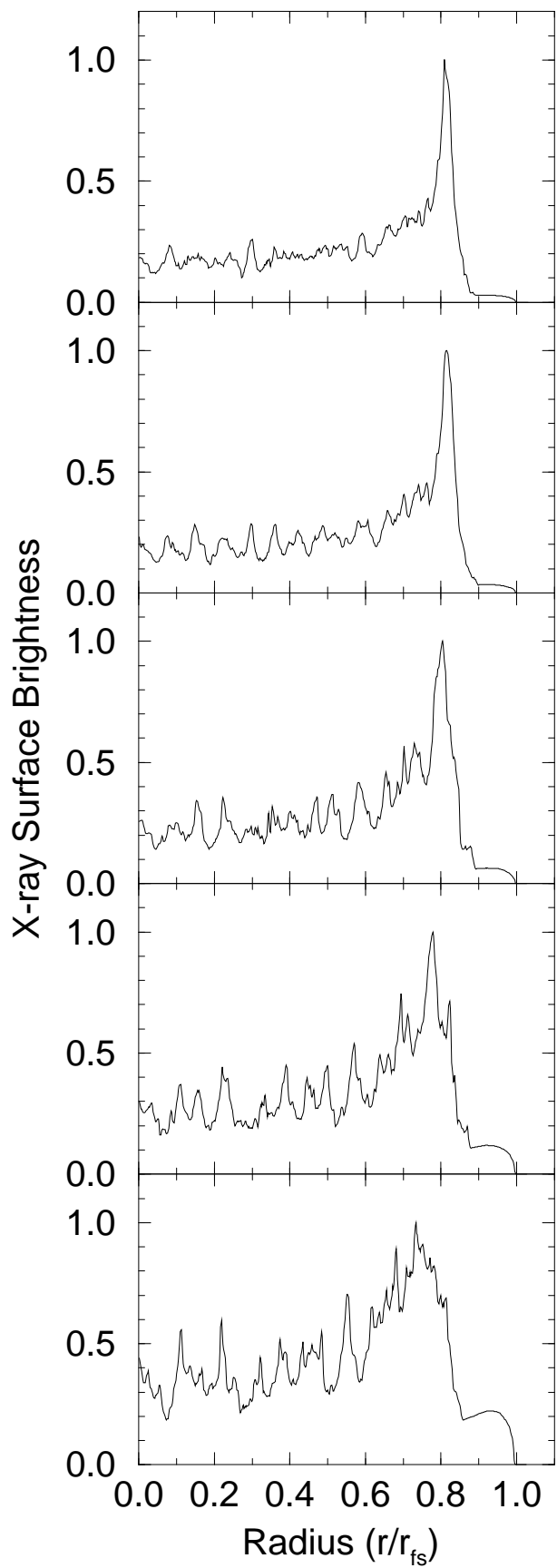
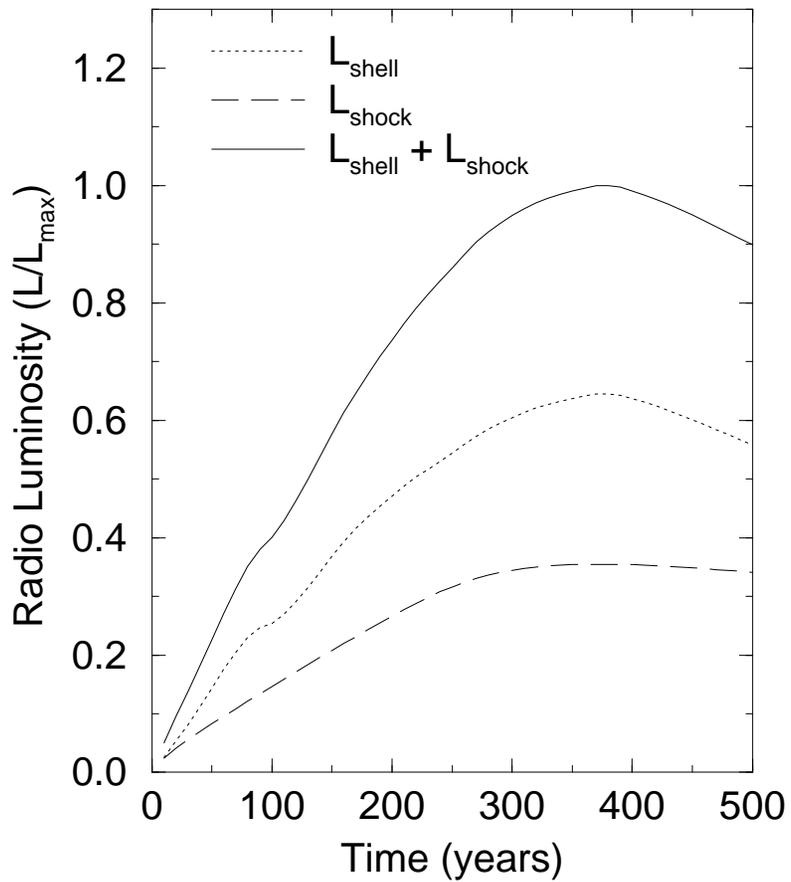
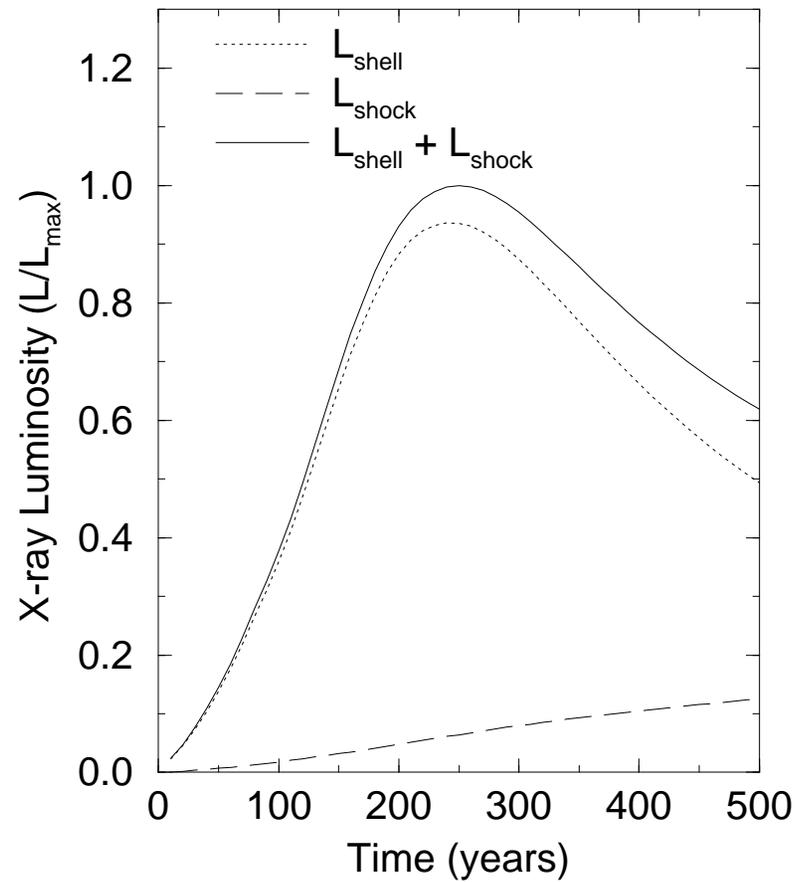
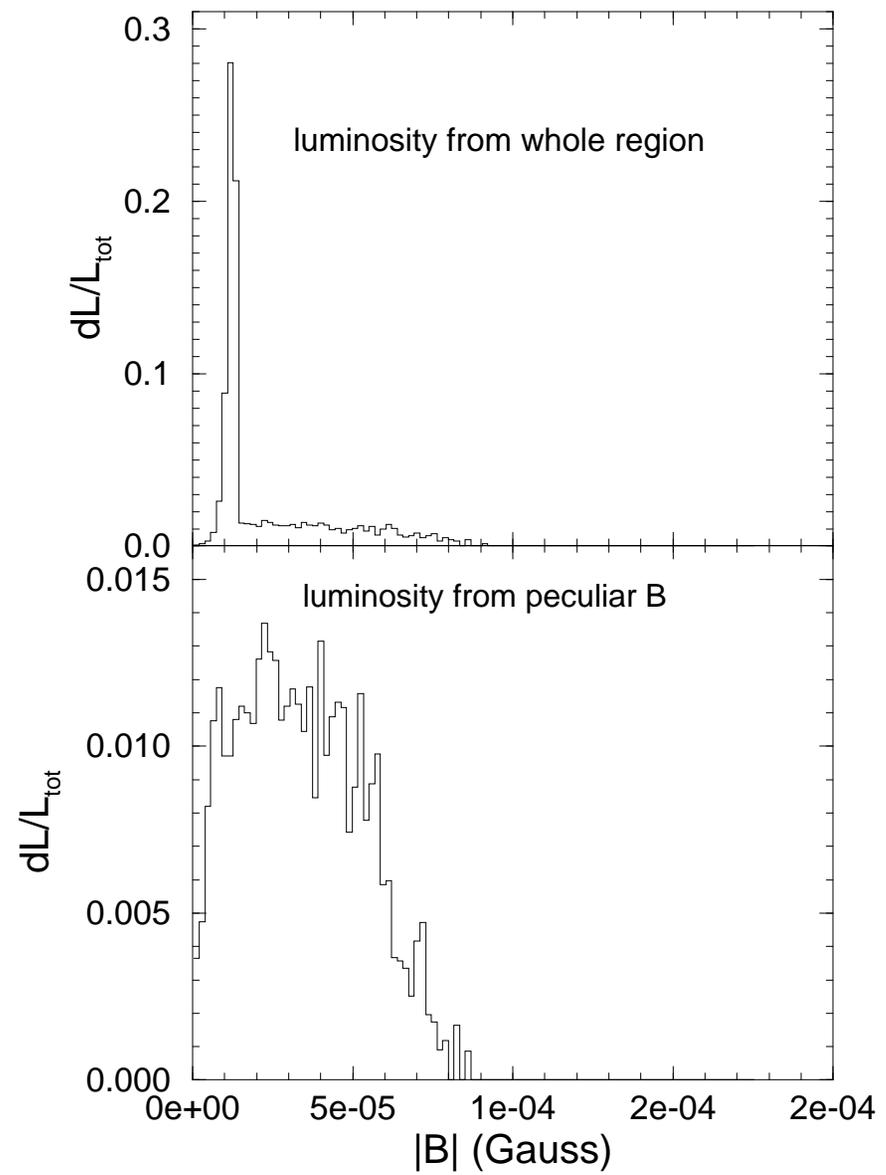
\end{document}